\documentclass[12pt,cite,epsf,epsfig]{article}
\usepackage{epsfig}
\title{PCT Theorem in Field Theory on Noncommutative Space}
\author{Namit Mahajan\thanks{E--mail : nmahajan@mri.ernet.in}\\
	{\em Harish-Chandra Research Institute,} \\
	 {\em Chhatnag Road, Jhunsi, Allahabad - 211019, India.}}

\def\be{\begin{equation}}
\def\ee{\end{equation}}
\def\bea{\begin{eqnarray}}
\def\eea{\end{eqnarray}}

\begin{document}
\maketitle

\begin{abstract}
The PCT theorem is shown to be valid in quantum field theory formulated on noncommutative spacetime by exploiting the
properties of the Wightman functions defined in such a set up.
\end{abstract} 

\begin{section}*{}
\indent Quantum field theory provides the general basis and correct theoretical framework known till date
 to describe particle properties and 
interactions and the PCT theorem enjoys a unique status in any viable and physically appealing
quantum theory. Stated crudely, it demands the invariance of the theory under the joint action (in any order) of the 
three discrete symmetries - parity (P), charge conjugation (C) and time reversal (T) and predicts the
existence of an anti-unitary operator realizing this joint symmetry. In an attempt to obtain the spin-statistics
connection, similar to that of Pauli \cite{paulispin}, Schwinger \cite{schwinger} assumed some form of PCT 
invariance. Without considering parity violation, it was noted by Luders \cite{luders} that charge conjugation
and time reversal impose the same restrictions on the theory i.e. if a relativistic quantum field theory
has space inversion then it must have the product of charge conjugation and time reversal as a symmetry. However,
it was realized by Pauli \cite{paulipct} that PCT itself is always a symmetry. There are two approaches to prove or
check the validity of this theorem:\\
(A1) It can be shown that the product of P, C, T taken in any order is same as Strong Reflection (SR) followed by Hermitian
conjugation. By this we imply the following set of conditions :\\
\bea \nonumber
x &\to& -x \\ 
O_{\mu_1\mu_2.....\mu_n}(x) &\to& (-1)^nO_{\mu_1\mu_2.....\mu_n}(-x) \\ \nonumber
\psi(x) &\to& i\gamma_5\psi(-x)
\eea
where $O_{\mu_1\mu_2.....\mu_n}(x)$ denotes a bosonic field carrying n number of tensorial indices. Given these transformation
rules, and given the fact that PCT is same as SR followed by hermitian conjugation, it becomes very easy to check
whether a given theory (with interactions) preserves PCT symmetry.\\
(A2) The second approach is to proceed via the axiomatic field theory route \cite{streater}. In this case, the whole theory can be 
reconstructed in terms of the Wightman functions defined as the vacuum expectation values of products of fields (generically 
denoted as $O_i(x_i)$ without worrying about the Lorentz indices)
\be
W(x_1,x_2,....,x_n) = \langle 0\vert O_1(x_1)O_2(x_2)....O_n(x_n)\vert 0\rangle \label{eq.1}
\ee 
satisfying very general conditions. \\ \\
\indent The basis of formulation of a quantum field theory, whether it is the Lagrange formalism or the axiomatic formalism, is 
a set of axioms. Without going into details of the proofs and generalities of the axioms, we quickly review 
the important features and summarise them as follows:\\
(Ax1) The states are described by unit rays in a physical Hilbert space and the state space possesses relativistic invariance.\\
(Ax2) The spectral condition implying that the physical four momenta lie in or on the positive light cone.\\
(Ax3) Existence of a unique vacuum.\\
(Ax4) Notion of a quantum field and its domain.\\
(Ax5) Poincare invariance of the fields i.e. the fields transform in a fixed manner under SL(2,C).\\
(Ax6) Local commutativity (sometimes also known as micro-causality) implying commutation (or anticommutation for fermions) of any
two field components for space-like regions.\\
(Ax7) The condition of asymptotic completeness i.e. demanding on physical grounds the validity of the following relation:
$~~{\mathcal{H}} = {\mathcal{H}}^{in} = {\mathcal{H}}^{out}~~$,
where $in$ and $out$ refer to the incoming and outgoing collision states.\\
It can be shown on general grounds that a quantum theory which satisfies all these axioms respects the PCT symmetry and the
normal spin-statistics relations.
The reason why PCT holds a very sacred place in any quantum field theory is that apart from being the outcome
of very general features in the theoretical formulation, till date no experiment has found deviations from the consequences
of this result, namely equality of mass and life times for particle and its anti-particle \cite{pdg}. In spite of these
reasons, this area has attracted a lot of attention (for an overview see \cite{kostelecky}).
It has been realized that the requirement of local commutativity (or locality)
 is too strong a condition and it should suffice to
prove the PCT theorem if this condition is relaxed to Weak Local Commutativity (WLC) which, for two fields, reads
\be
\langle 0\vert [O_1(x),O_2(y)]\vert 0 \rangle = 0 \hskip 1cm (x~-~y)^2 < 0
\ee
for any two bosonic operators (and the commutator $\to$ anticommutator for fermions). Therefore in the modified form, the PCT theorem
can be stated as the equivalence of weak local commutativity to the existence of a PCT operator satisfying the usual
properties and leaving the vacuum invariant.  In general WLC implies (modulo a sign factor arising due to and depending on the
number of fermions permuted)
\be
\langle 0\vert O_1(x_1)O_2(x_2).....O_n(x_n)\vert 0 \rangle = \langle 0\vert O_n(x_n)....O_2(x_2)O_1(x_1)\vert 0 \rangle
\ee 
if the set $\{x_i\}$ is a Jost point meaning thereby that 
$(\sum_i \lambda_i \xi_i)^2 < 0$
for real $\xi_i = x_i-x_{i+1}$ and $\lambda_i$ being set of real non-negative numbers not all zero. \\ \\
\indent Our usual notions of the space-time being described by a suitable manifold and the points on it being
labelled by a finite number of real coordinates may not be completely correct at smaller and smaller distances 
(or large enough energies), implying that the assumption of space-time being a continuum may not be valid at
all scales. If that is true then the underlying theory has an intrinsic length scale involved which is usually 
associated with the Planck length. In fact, with the aim to circumvent the problem of infinities in quantum
field theories, Snyder \cite{snyder} showed that there exists a Lorentz invariant space-time with a natural unit of length.
The consequences of such a solution are the modifications in the commutation relations for the operators corresponding
to both coordinates and momenta. However, it is easily seen that such modifications only show up at extremely large 
energy/momentum scales and the low energy physics is well described by the ordinary quantum theory.
Motivated by some recent string theory arguments, the field theory formulation on the noncommutative spaces has 
attracted a lot of attention. For a review of field theories on noncommutative spaces and various related issues
see \cite{douglas}. In a noncommutative set up the usual notion of coordinates being commutative is given up and
the hermitian coordinate operators are assumed to satisfy the following commutation relation
\be
[{\hat{x}}^{\mu},{\hat{x}}^{\nu}] = i\Theta^{\mu\nu}
\ee
where $\Theta^{\mu\nu}$ is a real antisymmetric matrix. In the quantum theory, this requirement translates
into the fact that the product of operators be replaced by the Weyl-Moyal product (also called star product,
 denoted by $*$). For a constant  $\Theta^{\mu\nu}$ this implies that for two operators, the star product is given by
\be
f(x)\ast g(x) = \Bigg[exp\Bigg(\frac{i}{2}\Theta^{\mu\nu}{\partial_{\eta}}_{\mu}{\partial_{\zeta}}_{\nu}\Bigg)f(x+\eta)
g(x+\zeta)\Bigg]_{\eta=\zeta = 0}   
\ee  
From the very expression it is obvious that theories on such spaces must be highly nonlocal due to the presence of
infinite number of derivatives. Also, such theories violate the Lorentz invariance in the sense of "particle Lorentz invariance"
\cite{kostelecky} while the observer Lorentz transformations remain the symmetry of the theory. \\ \\
\indent It is instructive to explore the validity of PCT theorem in a noncommutative theory where the locality is lost
from the very beginning. However, it seems natural to assume the validity of other axioms. In particular, since we would like
the theory to be describing physical situations, axioms Ax1-Ax5 should still hold and this is a reasonable assumption to 
start with. Also, the requirement of asymptotic completeness can be demanded on physical grounds. Therefore, but for the locality
condition, other axioms are satisfied and it is thus left to check whether with this condition not met, can PCT still hold.
Using the first approach (A1), the authors of \cite{chaichian} have shown that PCT theorem is valid for any form
of noncommutativity.\footnote{For noncommutative QED the discrete symmetries and their joint action has been discussed
in \cite{jabbari}. Using the arguments based on Lorentz violation, the authors of \cite{carroll} conclude that PCT is 
preserved in any realistic theory. The C, P and T transformation properties and hermiticity of the Seiberg-Witten maps
has been discussed in \cite{wess}.}
 In the present note, we would like to follow the other approach (A2) and see whether the same conclusions
can be reached by employing Wightman functions. \\ \\
\indent In analogy with the Wightman function defined in the ordinary theory Eq.(\ref{eq.1}), we define the
Wightman function in the noncommutative theory as follows
\be
W(x_1,x_2,....,x_n;*) = \langle 0\vert O_1(x_1)O_2(x_2)....O_n(x_n)\vert 0\rangle_* \label{eq.2}
\ee 
where $*$ refers to the product of fields defined appropriately in the noncommutative theory.
 In our calculations we will work only with scalar fields. Generalization
to fermions or vector particles is straight forward and the basic essence is the same. In the ordinary theory the commutator
of two scalar fields vanishes for space-like separations (locality implies weak local commutativity and thus PCT).
 This is clearly not the case here. Therefore, the main task is to examine the weak local commutativity condition and
see what we get out of it. Before exploring the general n point Wightman function, we take a close look
at the two point Wightman function defined as
\be
W_{AB}(x,y;*) \equiv \langle 0\vert\phi_A(x)\phi_B(y)\vert 0 \rangle_*  \label{eq.3}
\ee   
and the expectation value of the commutator
\be 
\langle 0\vert [\phi_A(x),\phi_B(y)]_*\vert 0 \rangle \label{eq.4} 
\ee
where again the subscript $*$ reminds of the fact that the products involved are the appropriate products
referred to as above. Eq.(\ref{eq.4}) is nothing
but the spectral representation. In the ordinary field theory it reads (for the same fields ie for A and B to be the same) 
\cite{itzykson}
\[
\langle 0\vert [\phi(x),\phi(y)]\vert 0 \rangle = i\int_0^{\infty}d{m^{\prime}}^2\sigma ({m^{\prime}}^2)\Delta (x-y;m^{\prime})
\]
where $\sigma(q^2)$ is related to the spectral density $\rho(q)$ in the usual way ($\rho(q)~=~\sigma(q^2)\theta(q^0)$ because
of Lorentz invariance and $\theta(q^0)$ is the step function)
and $i\Delta (x-y;{m^{\prime}})$ is the free
field commutator which vanishes outside the light cone. However, due to presence of noncommutativity the symmetry make be more
restricted and therefore in the noncommutative case we expect the noncommutative parameter to show up. On general grounds
it can be argued that the commutator in the noncommutative case should have the following spectral representation
\be
\langle 0\vert [\phi(x),\phi(y)]_*\vert 0 \rangle =  i\int_0^{\infty}d{m^{\prime}}^2\sigma ({m^{\prime}}^2,
i{\tilde{\partial}}^2)\Delta (\xi;m^{\prime}) \label{eq.5}
\ee
where ${\tilde{\partial}}^{\mu}~=~\Theta^{\mu\nu}\partial_{\nu}$ and $\xi~=~x-y$. Being able to write in this form is the
consequence of translational invariance which is not lost as $\Theta$ is independent of space-time variables. This is same
as the result obtained in \cite{liao}. Because
of the presence of derivatives in Eq.(\ref{eq.5}) it is not very clear whether the commutator expectation value will
vanish outside the light cone or not. However, it does vanish outside the light cone and the reason is not hard to see.
The derivatives (occurring as ${\tilde{\partial}}^2$) act only on $\Delta (\xi;m^{\prime})$ and this being a simple function
of exponentials gives back the same function with multiplicative factors proportional to ${\tilde{p}}^2$. For the n-th
term we will typically have
\be
({\tilde{\partial}}^2)^n\Delta (\xi;m^{\prime}) \propto  ({\tilde{p}}^2)^n\Delta (\xi;m^{\prime})
\ee
thus still preserving the space-time dependence of $\Delta (\xi;m^{\prime})$. The net result of all such terms can be
put in a series and we have the final form
\be
\langle 0\vert [\phi(x),\phi(y)]_*\vert 0 \rangle \sim \sum_{n=0}^{\infty}\frac{(-i{\tilde{p}}^2)^n}{n!}
 i\int_0^{\infty}d{m^{\prime}}^2\sigma ({m^{\prime}}^2)\Delta (\xi;m^{\prime}) \label{eq.6}
\ee
Therefore, even in the noncommutative case, the right hand side vanishes for $\xi^2~<~0$.\footnote{In presenting this
argument we have implicitly assumed that the power series has a smooth convergence, thereby allowing analytic continuation
to the desired domain. However, if this requirement is not met, then the arguments would have to be correspondingly
modified to allow for convergence, if possible, in an appropriately defined limit.}
 We therefore have the result that
WLC holds for the field theory formulated on noncommutative space and therefore this result can be extended to non-identical
fields and an arbitrary number of them. Specifically for the two point Wightman function we thus have the result
$W_{AB}(x,y;*) = W_{BA}(y,x;*)$ for space-like separation between them. Therefore WLC imples
\be
W_{AB}(\xi;*) = W_{BA}(-\xi;*) \hskip 1cm ~for~~~ \xi^2<0
\ee
We emphasize again that the symbol $*$ in the above equations and expressions should be inferred on the lines of Eq.(\ref{eq.5})
and Eq.(\ref{eq.6}) and should not be confused with the $*$ appearing in the formal expression of the Moyal product. It is just to
represent the quantities in noncommutative field theory and distinguish them from their counterparts in the ordinary theory and
all the information related to the fact that we are now dealing with a quantum theory on a noncommutative space is coded in
the suitably defined spectral density and similar objects of interest.\\ \\
\indent Also, if $\xi$ is space-like then under a Lorentz transformation $\xi \to -\xi~~$, $W_{AB}(\xi;*) ~=~W_{AB}(-\xi;*)$. Consider now the
following quantity:
\be
W_{[AB]}(\xi;*) = W_{AB}(\xi;*) - W_{BA}(\xi;*)
\ee
Clearly for space-like $\xi$, $~W_{[AB]}(\xi;*)$ vanishes if WLC holds. But the two point Wightman function and the permuted one
 can both be defined as the boundary values of holomorphic functions in a complex plane. Therefore, the vanishing of 
$W_{[AB]}(\xi;*)$ for space-like $\xi$ on the real axis implies that $W_{[AB]}(\xi;*)~=~0$ everywhere.
Therefore, we have $W_{AB}(\xi;*)~=~ W_{BA}(\xi;*)$ which is same as SR in terms of Wightman functions.
The same analysis can be extended for fermions where the commutators will be replaced by anti-commutators.
For the case of n point Wightman function, the steps remain essentially the same.\\ \\
\indent Noncommutative quatum field theories are afflicted by the phenomenon of UV/IR mixing and related
pathologies \cite{shiraz}. The presence of hard divergences due to this effect may spoil the very basis of Wightman formalism completely,
rendering no or very little scope for operations like analytic continuation and smooth convergence of power 
series etc. In order that the quantum theory makes sense physically, such divergences should not manifest themselves
in any form. We assume that some mechanism which by-passes the difficulties arising due to UV/IR mixing
is at play and that UV/IR mixing poses no serious threat to the results obtained.
For the present case it is not required and therefore we do not bother about the details
of such an underlying mechanism.\\ \\
\indent We briefly comment on the types of noncommutativities and possible difficulties relating to them.
For space-space noncommutativity, since the star product does not involve
time derivatives, there are no subtelties involved in handling the time ordered products. 
Therefore, it is easier to see that the theories with only space-space noncommutativity
preserve the gross features and axioms of the quantum theory to a greater extent. It becomes far more
complicated and involved if space-time noncommutativity is present. A careful handling of time ordering
has to be employed because of the presence of an infinite number of time derivatives which can lead to
very complicated and non-compact looking results and expressions. To interpret these results would require 
further care. Moreover, simplistic and naive treatment should be abandoned in order to obtain sensible results.
With time being involved in the noncommutativity, more conceptual issues at the level of foundations of 
quatum theory may creep in and would require a detailed and more careful treatment. However, for the present case,
 we tend to ignore all such issues and base our arguments on the hope that such issues don't hamper
the basic axioms of the quantum field theory that we have employed, thereby allowing us to reach the
desired conclusions. An important thing to remember is the fact that the symmetry is much more restricted in
the present situation and depending upon the type of noncommutativity, the theory admits a particular symmetry group
and associated structure. Therefore, a careful analysis, keeping track of and correctly taking into account the
nature and extent of the allowed operations in such a case like analytic continuation etc, should yield these results.\\ \\  
\indent We therefore see that by looking at the behaviour of Wightman functions it is clear that PCT theorem is
valid in quantum field theory defined on noncommutative space, provided we assume the validity
of axioms Ax1-Ax5. Nowhere in the whole analysis has any specific form of
noncommutativity assumed, except for the assumption that $\Theta$ is independent of space-time variables. As in
\cite{chaichian}, it remains to be explored in detail whether some specific form of noncommutativity leads to violation
of spin-statistic relation in context of Wightman function approach as well. Also, the more conceptual and philosophical
issues concerning the time direction being noncommutative have still to be explored and investigated in detail.
However, it is very clear that the results obtained hold without much doubt 
if we restrict ourselves to space like noncommutativity.\footnote{While this work was completed a similar work
\cite{alvarez} appeared. The results more or less match, atleast in the domain of validity of the axioms that 
we have based our arguments on.}\\ \\ 
{\bf Acknowledgements} I would like to thank E.~Harikumar and Kamal Datta for their comments.

\end{section}

%
\end{document}